**Title: Universal Kinetics of the Sol-to-Gel Transition**

**Authors:**

Pai Liu[1,*], William R. Heinson[1,*], Christopher M. Sorensen[2] and Rajan K. Chakrabarty[1,3,†]

**Affiliations:**

[1] Center for Aerosol Science and Engineering, Department of Energy, Environmental and Chemical Engineering, Washington University in St. Louis, St. Louis, Missouri 63130, USA

[2] Department of Physics, Kansas State University, Manhattan, Kansas 66506, USA

[3] McDonnell Center for the Space Sciences, Washington University in St. Louis, St. Louis, Missouri 63130, USA

[*] These authors contributed equally to this work

[†] Corresponding author: chakrabarty@wustl.edu

**ABSTRACT**

A comprehensive theory encompassing the kinetics of sol-to-gel transition is yet to be formulated due to break-down of the mean-field Smoluchowski Equation. Using high temporal-resolution Monte Carlo simulation of irreversible aggregation systems, we show that this transition has three distinct regimes with kinetic exponent $z \in [1, 2)$ corresponding to aggregation of sol clusters proceeding to the ideal gel point (IGP); $z \in [2, 5.7)$ for gelation of sol clusters beyond IGP; and $z \in [2, 3.5)$ for a hitherto unidentified regime involving aggregation of gels when monomer-dense. We further establish universal power-law scaling relationships that connect the kinetics of these three regimes.

Aggregation is a phenomenon ubiquitous in colloidal and aerosol systems [1-15]. Upon dispersion, particles collide and often irreversibly stick together to form larger clusters. Provided the monomer-monomer contact is non-coalescent in nature, the aggregates manifest a scale-invariant, fractal-like morphology quantifiable with a mass fractal dimension ($D_f$) [1-3]. Prolonged aggregation leads to the phenomenon of gelation – a process involving the jamming together of ramified aggregates and the formation of volume-spanning networks with a characteristic $D_f \approx 2.5$ [4-15]. Gelation, as a phenomenon, has opened many avenues for synthesis of materials with unique properties [5, 16-19]. The contemporary application of gelation theory extends to a broader context, for example, predicting the influence of wild fire emissions on climate change [20-22] and counteracting the formation of online extremists group supports [23]. Despite its wide-ranging applicability, the theory of gelation still grapples with the fundamental questions "*How fast does a sol system gel and what are the associated critical time scales?*" The difficulty in formulating a comprehensive kinetic theory stems from the break-down, at the onset of gelation, of the mean-field assumption which lies at the core of the governing Smoluchowski Equation (SE) [7-9, 15, 16, 24]. That is, SE, which tracks the time evolution of the system can only go so far in predicting the pre-onset of gelation, but not gelation itself [7, 8]. An alternate albeit successful interpretation of the sol-gel phenomenon is the percolation theory, which is a static model and hence, cannot predict the kinetics [15]. In this Letter, we address this long-standing problem by establishing a set of system-independent kinetic expressions capable of predicting the complete evolution of the sol-to-gel process. We do so by performing high temporal-resolution analysis of the evolution of diffusion limited cluster-cluster aggregation (DLCA) systems, which have been recently shown to produce gels that share identical morphologies with those produced via the percolation model [15].

The tendency of gelation stems from the simple fact that the $D_f$ of non-coalescent aggregates is always smaller than the spatial dimension ($d = 3$) in three-dimensional space [7, 8]. When aggregates grow with $D_f < d$, the increase in their average size outruns their average nearest-neighbor separation. That said, the system inevitably evolves to crowded states [7, 8, 16]. This crowding persists until ultimately all the free-space in the system is occupied, at which point the system is said to reach the ideal gel point (IGP) [7, 8, 16]. Subsequently, the sol-to-gel transition occurs, and when all aggregates in the system are incorporated into one single volume-spanning particle, the system reaches the final gel state (FGS) [7]. The existing kinetic theory discusses the sol-to-gel transition within two separate regimes per the applicability of SE [7, 8, 15, 16]. Prior to IGP, the evolution of sol systems could be described as binary, mean-field, cluster-cluster aggregation of particles (hereafter Regime I) with kinetics governed by SE [7, 8, 15, 16]. Solution to SE with homogeneous kernel leads to the scaling relationship between total number of clusters in the system ($n_{tot}$) and inverse time ($t^{-1}$), $n_{tot} \propto t^{-z}$, where the kinetic exponent ($z$) quantitatively measures how fast aggregation proceeds [7, 8, 24, 25]. For example, when a system starts out dilute, Brownian aggregation mechanism holds, which yields $z = 1$ [7, 8, 24, 25]. Past studies have demonstrated that the kinetics of aggregation tend to speed up as the system evolves to crowded states and $z$ continuously increases to about two when the systems reach IGP [7, 8]. Such enhancement in kinetics has also been observed to continue beyond IGP, where gelation of sol clusters (hereafter, Regime II) takes precedence. In this regime, the enhanced kinetics have been reported by Fry et al. (2002) who empirically extrapolated and mapped the values of increasing $z$ with respect to $t$ [7, 8].

Here, we systematically study the late-stage kinetics by numerically simulating the gelation system using the off-lattice DLCA model [15, 26, 27]. The algorithm starts out by generating a cubic simulation box with three-million randomly placed monomers. The monomer volume fraction ($f_{vm}$) is controlled by specifying the volume ($V$) of the simulation box,

$$f_{vm} = \frac{4}{3}\pi a^3 \frac{n_{tot,0}}{V} \qquad (1)$$

where $a$ is a monomer radius in arbitrary units, and $n_{tot,0}$ denotes the total number of particle (cluster) at $t = 0$, which is equal to the conserved total number of monomers in the system. The simulation proceeds by randomly picking a cluster of mass $N$ (number of constituent monomers, and $N = 1$ for monomer) and moving it by $2a$ in a random direction with probability $N^{-1/D_f}$ per Stokes-Einstein diffusion [24]. The algorithm tracks the total number of clusters ($n_{tot}$), and once $n_{tot}$ clusters are picked, $t$ is incremented by unit simulation time $t_s$. We define $t_s$ as the time-interval during which monomers move by a root-mean-squared-displacement of $2a$:

$$t_s = \frac{4\pi\mu a^3}{k_B T} \qquad (2)$$

where $k_B$, $T$, and $\mu$, respectively represent the Boltzmann constant, temperature, and viscosity of the surrounding gas (See Supplementary Sect. I for derivation) [24]. During the process, if two clusters collide, they are joined together forming a new cluster, and $n_{tot}$ decreases by one. The algorithm repeats the above procedure until $n_{tot} = 1$, that is, the FGS is attained, and the corresponding time $t_{FGS}$ is recorded. Next, we study the evolution of the system by analyzing the mass frequency distribution (hereafter, mass distribution) of clusters, $n(\log_{10}N, t/t_s)/n_{tot}(t/t_s)$, where $n$ denotes the number of clusters (having $N$ monomers at $t/t_s$).

FIG. 1 shows the aggregate mass distributions in the $\log_{10}N - t/t_s$ space for systems of various $f_{vm}$. Panel (a) demonstrates the mean-field growth of sol clusters typically seen in transition Regime I, during which the kinetics could be described with the exact solution to SE [7, 8, 24, 25]. Next, we discuss the onset of gelation and the subsequent Regime II by comparing the cluster mass distributions with the analytical solution values of average cluster mass at the IGP ($N_{IGP}$):

$$N_{IGP} = k_0(R_{g,IGP}/a)^{D_f} \qquad (3.1)$$

$$\text{with } R_{g,IGP} = a\left[f_{vm}^{-1}k_0\left(\frac{D_f}{D_f+2}\right)^{3/2}\right]^{1/(3-D_f)} \qquad (3.2)$$

where $R_{g,IGP}$ is the average linear size (radius of gyration) of aggregates at IGP and $k_0$ is a fractal prefactor. Eq. (3.1) follows mass scaling power-law relationship with $k_0 = 1.3$ and $D_f = 1.8$ describing the morphology of DLCA. Eq. (3.2), originally introduced in Ref [12], is reached when one equalizes the system $V$ to the total cluster perimeter volume (volume of the minimum encapsulating spheres, hereafter MES). FIG. 1 (b) – (e) show that when sol clusters grow, their geometric mean mass $\langle \log_{10}N \rangle$ (red dashed lines) asymptotes to the $\log_{10}N_{IGP}$ value (red solid lines) predicted by Eq. (3). Subsequently, the IGP could be identified at the point where $\langle \log_{10}N \rangle$ reaches $\log_{10}N_{IGP}$ (See Supplementary Sect. II), and the corresponding time is regarded as the ideal gel time ($t_{IGP}$). In FIG. 1 we mark the IGP at the points ($N = N_{IGP}, t = t_{IGP}$) using triangle notes. One could observe that immediately after IGP, the mass distribution (in FIG. 1(b) – (e)) develops into a bimodal one, indicating the onset of a separate phase, the gel. Subsequently, the gel clusters in the systems continuously grow by scavenging the remaining sol clusters. We mark the FGS in FIG. 1 at the points ($N = n_{tot,0}, t = t_{FGS}$) using circle notes. Comparing the characteristic timescales $t_{IGP}$ and $t_{FGS}$, one could observe that beyond IGP, a considerable amount of time is needed for the total conversion of sol clusters to gel phase, during which the mass distribution of sol clusters stays invariant with $\langle \log_{10}N \rangle$ closely matching $\log_{10}N_{IGP}$. Such invariance indicates that beyond IGP, there is no more free space for sol clusters to grow any further without turning into gel phase.

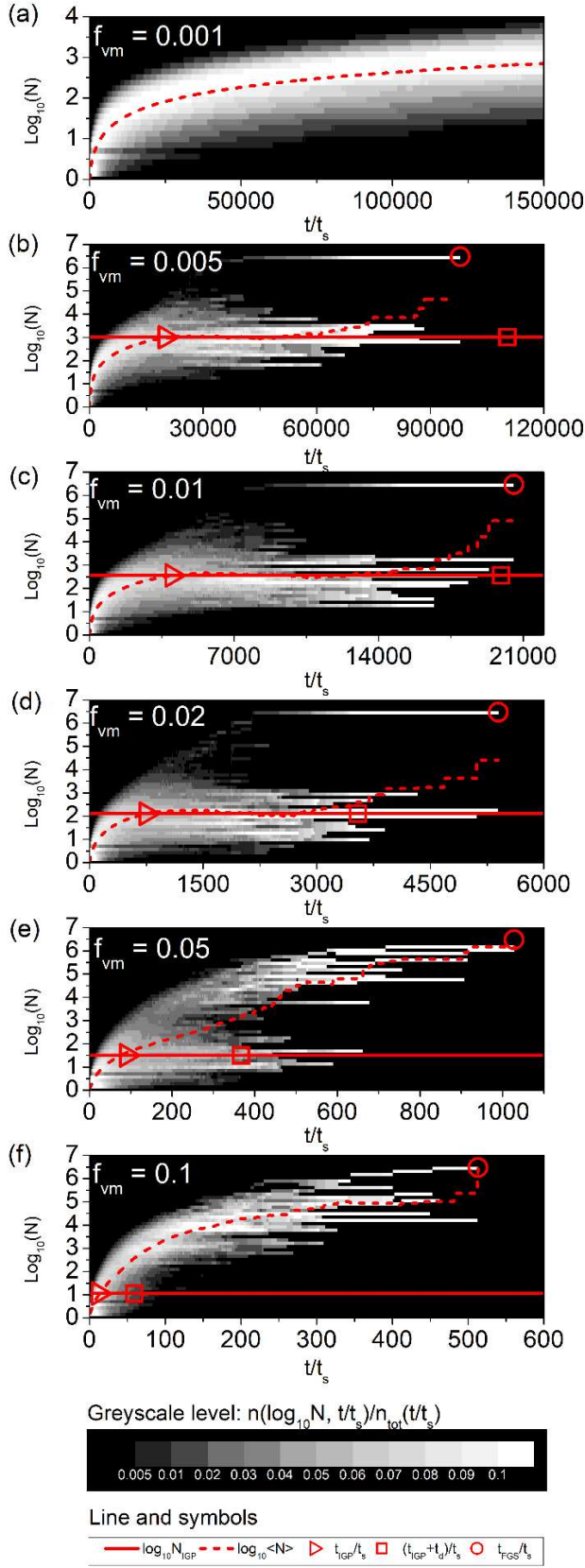

FIG. 1. Time-evolution of the aggregate mass distribution for DLCA systems with $f_{vm}$ = 0.001 (a), 0.005 (b), 0.01 (c), 0.02 (d), 0.05 (e), and 0.1 (f). The solid lines represent the analytical solution values to the characteristic cluster mass at IGP, $\log_{10} N_{\text{IGP}}$, which follows Eq. (3). The dashed lines represent the geometric mean values of cluster mass, $\langle \log_{10} N \rangle$. In panel (b)-(f), triangles and circles respectively represent $t_{IGP}$ and $t_{FGS}$, which are determined from the simulations. Squares predict the time when sol clusters deplete, $t_{IGP} + t_d$, and $t_d$ is solved using Eq. (4). All timescales are presented in units of $t_s$

We next quantify the length of the time interval between IGP and the total conversion of sol clusters, which is defined as $t_d$ in the context of this Letter. We start by asking why such latency period $t_d$ would exist? Clusters, which are fractals, only fill the Euclidean space partially. As a result, IGP (at which the MES of sol clusters saturate system volume) does not guarantee that all sol clusters have encountered their neighbors and hence established connections. Following this logic, we write $t_d = \frac{(V_{MES,IGP} - V_{agg,IGP})^{2/3}}{6 D_{agg,IGP}}$, where $V_{agg,IGP}$ is the volume of the sol cluster with mass of $N_{IGP}$; $V_{MES,IGP}$ denotes the volume of the corresponding MES; and $D_{agg,IGP}$ represents the diffusivity of that cluster. The intuition behind this scaling analysis is that beyond the IGP the probability for a sol cluster to be scavenged reaches unity, if that cluster has thoroughly explored all the free volume enveloped within its MES, and till then $t_d$ has elapsed from $t_{IGP}$. We show a step-by-step derivation in Supplementary Sect. III that $t_d$ after being normalized by $t_s$ is a function of $f_{vm}$ only, written as:

$$\frac{t_d}{t_s} = \frac{1}{4}\left[\frac{4}{3}\pi(1-f_{vm})\right]^{2/3} \left(\frac{D_f}{D_f+2}\right)^{(2D_f+3)/(6-2D_f)} (k_0 f_{vm}^{-1})^{3/(3-D_f)} \qquad (4)$$

In FIG. 1 we mark the critical points corresponding to the total conversion of sol clusters at $(N = N_{IGP}, t = t_{IGP} + t_d)$ using square notes. Good agreements between the predictions by Eq. (4) and the simulations are seen for systems with changing $f_{vm}$. Another important inference made here is that $t_d$ is inversely related to the $f_{vm}$. Qualitatively, this is because a denser (higher $f_{vm}$) system reaches IGP with average sol clusters of smaller $N_{IGP}$, which have smaller $V_{MES,IGP} - V_{agg,IGP}$, but larger $D_{agg,IGP}$. One could also observe such inverse correlation between $t_d$ and $f_{vm}$ in FIG. 1(b) – (e). For example, when a system is sufficiently dilute ($f_{vm} = 0.005$ and $0.01$), the total conversion of sol clusters occurs at a timescale comparable to that of the FGS, formally written as $t_{IGP} + t_d \approx t_{FGS}$. With a further increase in $f_{vm}$ (from 0.02 to 0.05), $t_d$ decreases, the total conversion of sol clusters precedes FGS, and the time interval between $t_{IGP} + t_d$ and $t_{FGS}$ becomes significant. Such tendency reaches an extremity when $f_{vm} = 0.1$, as shown in FIG. 1(f). One could observe that $t_{IGP} + t_d \ll t_{FGS}$ and the system evolves with a unimodal cluster mass distribution throughout the entire process. This unimodality implies that sol clusters and gels no longer co-exist. For the extremely dense system ($f_{vm} = 0.1$), the transition process beyond the timescale of $t_{IGP} + t_d$ is defined as Regime III, which differs fundamentally from the classical view on the gelation of sol clusters [7, 8, 24, 25].

We next demonstrate that the kinetics corresponding to Regimes I, II and III could be unified on coherent power-law relationships, when the transition is observed with two timescales, first, the characteristic time for Brownian aggregation ($t_c$), and second, the $t_{IGP}$. In Regime I, solving SE with homogeneous Brownian kernel provides the scaling law with $z = 1$ [7, 8, 24, 25]:

$$\frac{n_{tot}}{n_{tot,0}} = \left(1 + \frac{t}{t_c}\right)^{-1} \tag{5}$$

$$\text{and, } t_c = \frac{3\mu V}{4k_B T n_{tot,0}} \tag{6}$$

According to Eq. (5), we empirically determine $t_c$ from DLCA simulations at the time when $n_{tot}$ decreases to half of the initial values (See Supplementary Sect. IV).

FIG. 2(a) shows that when $t$ is normalized per $1 + t/t_c$, the early stages of the transition are unified and the trends of $n_{tot}/n_{tot,0}$ follow Eq. (5) with $z = 1$, indicating the dominance of Brownian aggregation mechanism. This is especially true for $f_{vm} = 0.001$, whereas the behavior becomes more rapid than Eq. (5) for progressively larger $f_{vm}$. This deviation from Eq. (5) indicates subsequent cluster-dense conditions, and the kinetics speed up with the kinetic exponent $z$ taking on values larger than unity, during which the driving mechanism of aggregation becomes ballistic-limited as the interdigitating aggregates have no more free space to diffuse [7, 8, 28].

FIG. 2(b) shows the late-stages of the transition are unified upon normalizing $t$ according to $1 + t/t_{IGP}$. A universal power-law relationship manifests as,

$$\frac{n_{tot}}{n_{tot,IGP}} = 2^{z_{FGS}} \left(1 + \frac{t}{t_{IGP}}\right)^{-z_{FGS}} \quad (7)$$

where $z_{FGS} \approx 5.7$ is the kinetic exponent reported at the FGS [8]. The $z$ takes on a terminal value because only when $t \gg t_{IGP}$ (that is, FGS) could Eq. (7) be reduced to $n_{tot} \propto t^{-z}$. The prefactor takes up the expression of $2^{z_{FGS}}$, satisfying the condition of $n_{tot} = n_{tot,IGP}$ when $t = t_{IGP}$. The two power-law relationships, Eq. (5) and the new Eq. (7) provide a complete description for the full so-to-gel transition within regimes I and II.

Regime III occurs in system $f_{vm} = 0.1$ with a counterintuitively slower kinetics. FIG. 2(c) shows that the decrease in $n_{tot}/n_{tot,0}$ for $f_{vm} = 0.1$ falls behind that for $f_{vm} = 0.05$. When observed with $1 + t/t_{IGP}$, a less steep decreasing trend of $n_{tot}/n_{tot,IGP}$ is seen (Panel (d)) when $f_{vm} = 0.1$. Eq. (7) still holds valid while a $z_{FGS} \approx 3.5$ fits the data best, per the red solid line in panel (d). This slower rate – in a denser system – could be due to the abundance gel clusters which are considerably less mobile. Qualitatively speaking, the extremely dense system facilitates almost an instantaneous gelation of sol clusters, but the resultant abundance of gels slows down the system progressing from IGP to FGS. FIG. 2(d) shows that the decreasing trend of $n_{tot}/n_{tot,IGP}$ for the system with $f_{vm} = 0.05$ originally follows Eq. (7) with $z_{FGS} \approx 5.7$ until reaching an inflection point indicated by the arrow in (d). Beyond the inflection, the trend asymptotes to the less steep one governed by $z_{FGS} \approx 3.5$. Note that this inflection occurs approximately at $t_{IGP} + t_d$ of the system (see FIG. 1(e)), indicating that a slowing down of kinetics is indeed a characteristic of the system in which only gel clusters exist. These dense gelation systems near $f_{vm} = 0.1$ are traditionally discussed using the static percolation [5-7,15], and here we emphasize that the kinetic aspect should not be overlooked.

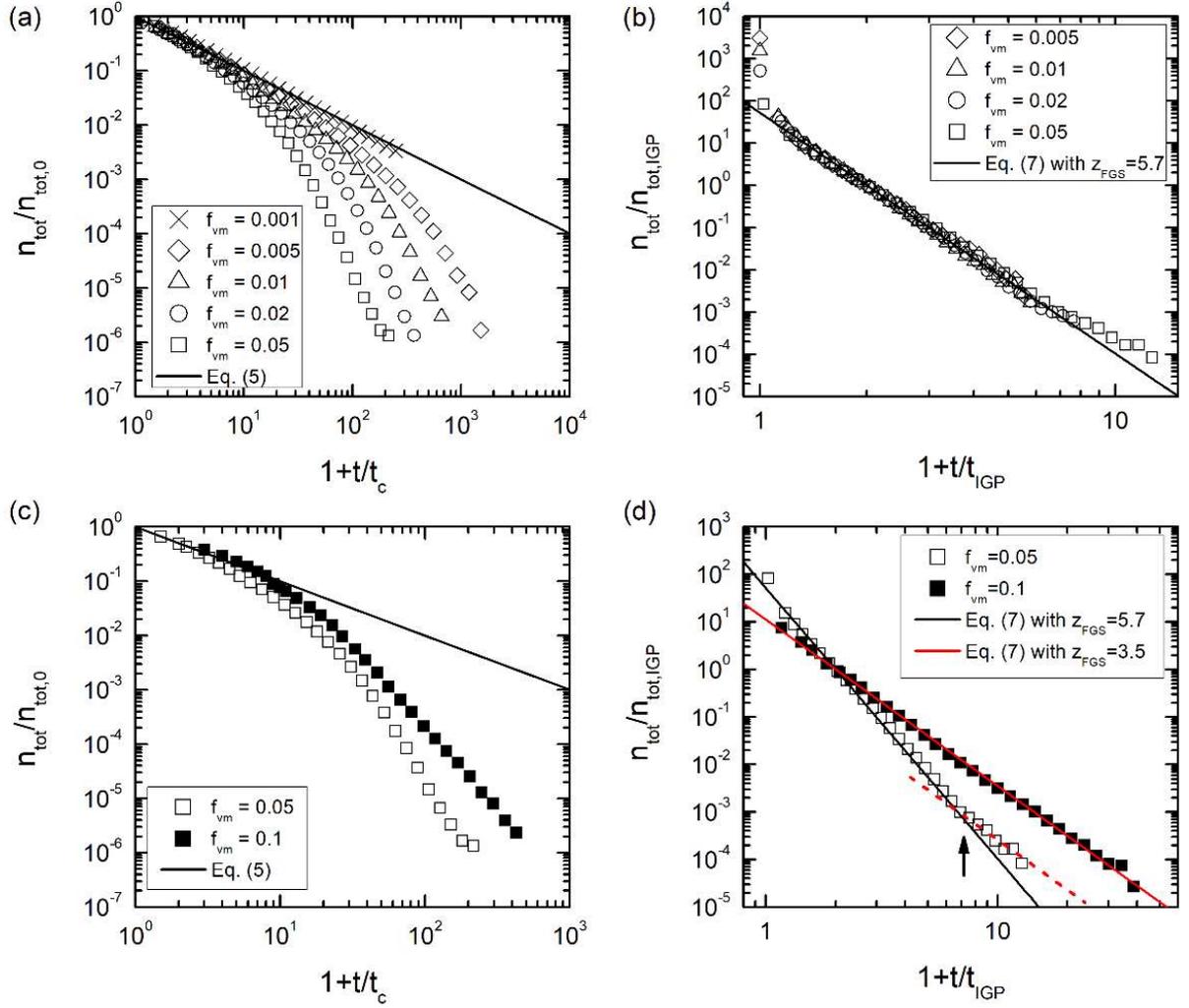

**FIG. 2.** (a) and (c) show that early stages of aggregation are unified on Eq. (5) when observed with normalized time $1 + t/t_c$. (b) and (d) show that the late stages of transitions, when observed with normalized time $1 + t/t_{IGP}$, are unified on Eq. (7). Dashed line in (d) has a slope of -3.5, and the arrow indicates an inflection in transition kinetics.

We next evaluate the existing analytical expressions for the characteristic timescales and provide improved parameterizations. Combining Eq. (6) with (1) and (2) yields the analytical expression for $t_c$ in units of $t_s$:

$$\frac{t_c}{t_s} = \frac{1}{4} f_{vm}^{-1} \quad (8)$$

FIG. 3 compares the prediction of Eq. (8) with the $t_c/t_s$ values determined from simulations. The exponent -1 fails at large $f_{vm}$, indicating that the aggregation in the extremely dense systems deviates from the Brownian mechanism at a very early point. The enhanced kinetics lead to the $t_c$ which are much smaller than what Eq. (8) predicts. The upper limit for an increasing $f_{vm}$ is the percolation threshold, at which point the volume-spanning gel instantaneously forms, and so that both $t_c$ and $t_{IGP}$ decrease to zero. Ref [29] reports the critical volume fraction ($\Phi_P$) to be about 0.18 (black line in FIG. 3) at which a system of randomly packed hard spheres reaches percolation threshold in three-dimensional space. We introduce a semi-empirical expression of $t_c$ for the dense DLCA systems near the percolation threshold as

$$\frac{t_c}{t_s} = \frac{1}{4}(f_{vm}^{-1} - \Phi_P^{-1}) \quad (9)$$

The prediction by Eq. (9) is plotted in FIG. 3 and it captures the dramatic decrease in the $t_c/t_s$ near $\Phi_P$. Similarly, we provide an improved parameterization of $t_{IGP}$ for these dense DLCA systems,

$$\frac{t_{IGP}}{t_s} = \frac{1}{4}(f_{vm}^{-1} - \Phi_P^{-1})\left[ b^{-\frac{1}{z}} f_{vm}^{\frac{D_f}{(D_f-3)z}} \left(\frac{D_f}{D_f+2}\right)^{\frac{3D_f}{(6-2D_f)z}} k_0^{\frac{3}{(3-D_f)z}} - 1 \right] \quad (10)$$

where $b$ originates from a power-law relationship $\langle N \rangle = b(1 + t/t_c)^z$ quantifying the aggregate growth when cluster-dense condition sets in [25]. The step-by-step derivation of Eq. (10) is outlined in Supplementary Sect. V. The determination of the values for $b$ and $z$ from DLCA is discussed in Sect. VI. FIG. 3 also compares the prediction of Eq. (10) (solved with $D_f = 1.8$, $k_0 = 1.3$, $z = 1.5$ and $b = 0.2$) to the $t_{IGP}/t_s$ determined from simulations. The traditionally used expression for $t_{IGP}$ [7, 16] is also evaluated here: $t_{IGP} \approx a^3 K^{-1} f_{vm}^{-2.5}$, which after being combined with Eq. (2) yields to:

$$\frac{t_{IGP}}{t_s} = \frac{3}{16\pi} f_{vm}^{-2.5} \tag{11}$$

FIG. 3 shows that Eq. (11) overestimates $t_{IGP}$ by a factor less than two. The summation of Eq. (10) (or (11)) and (4) provide analytical solution the values for $(t_{IGP} + t_d)/t_s$, which are compared with the $t_{FGS}/t_s$ determined from simulations. We again observe that a dilute system reaches FGS when the total conversion of sol clusters is attained ($t_{IGP} + t_d \approx t_{FGS}$), but when monomer dense, the time interval between $t_{IGP} + t_d$ and $t_{FGS}$ becomes significant, during which regime III takes over the kinetics.

We conclude this Letter with FIG. 4 which schematically illustrates the comprehensive picture of the full sol-to-gel transition. The transition Regimes I-III and the corresponding kinetic expressions (Eqs. (5) and (7)) are presented along with the characteristic timescales, $t_c$, $t_{IGP}$, and $t_{IGP} + t_d$ which serve as milestones. Please note that the regime I and II are separated per $t_{IGP}$ at which point SE breaks down, but kinetics 1 fails at a timescale $\approx t_c$ beyond which Brownian aggregation mechanism no longer holds valid [7, 8, 9, 15]. Regime II and III start out simultaneously at IGP, but II tends to dominate over III because number of gel clusters in a system is typically negligible compared to that of the sol clusters. Regime III only take precedence after II reaches its completion, that is $t_{IGP} + t_d < t < t_{FGS}$, which only manifests when monomer-dense.

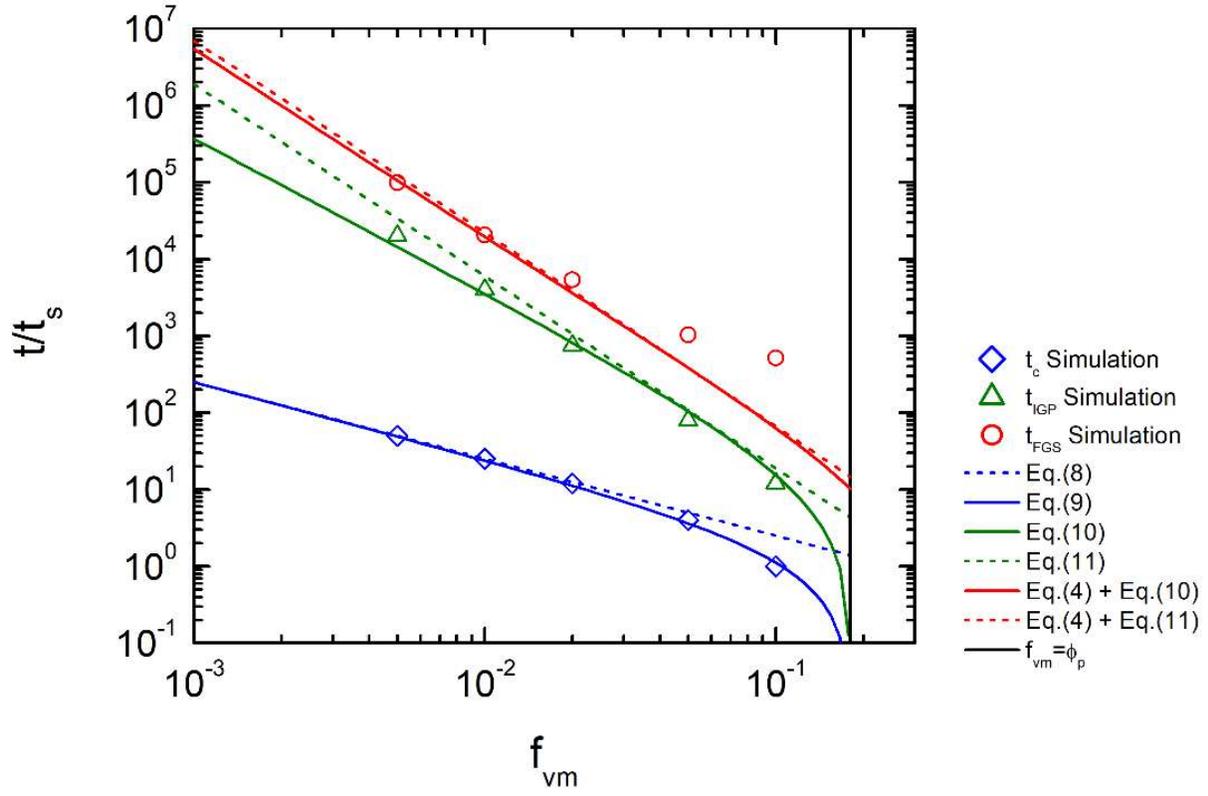

FIG. 3. Characteristic timescales $t_c$, $t_{IGP}$ and $t_{FGS}$ as functions of $f_{vm}$. The timescale parameters determined from DLCA simulations are compared with their analytical solution values. All timescales are reported in unit of $t_s$

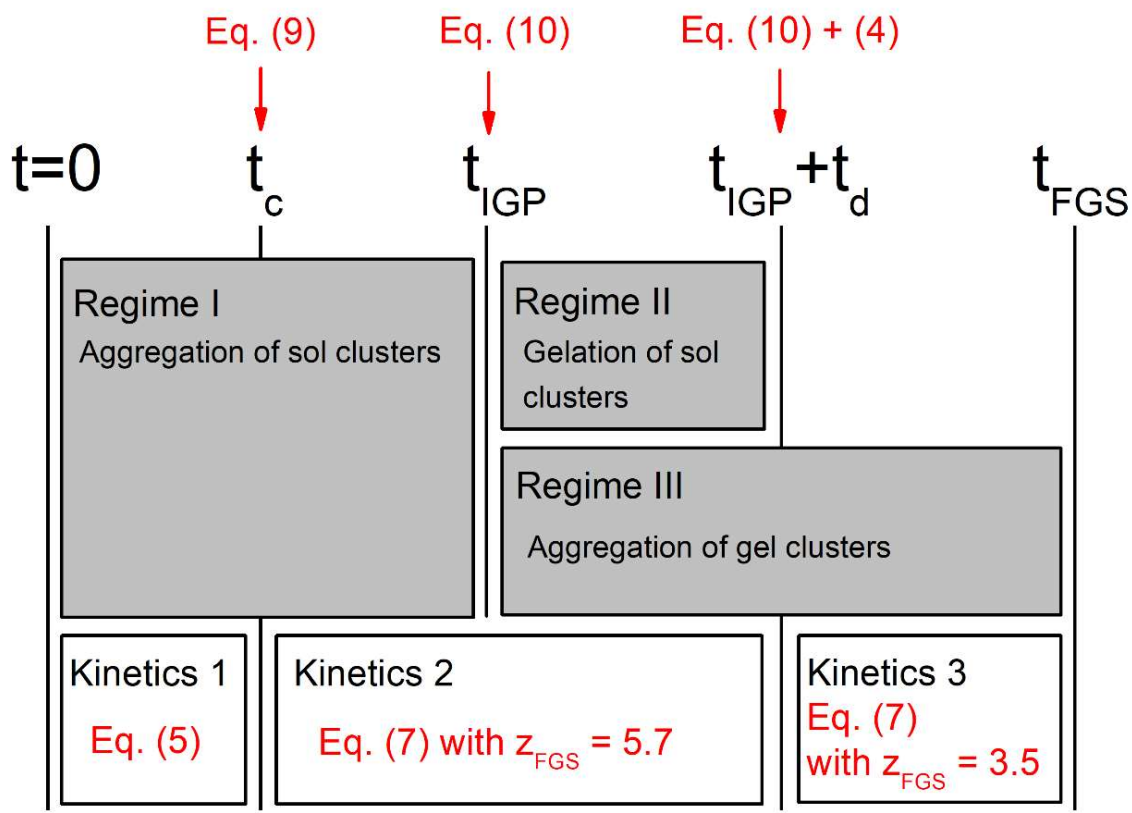

**FIG. 4. The kinetics of full sol-to-gel transition.**

ACKNOWLEDGEMENTS

This work has been supported by the US National Science Foundation (NSF) grants AGS-1455215, CBET-1511964, AGS-1649783 and AGS-1623808, and the NASA Radiation Sciences Program grant NNX15AI66G.

**Title: Universal Kinetics of the Sol-to-Gel Transition**

**Authors:**

Pai Liu[1, *], William R. Heinson[1, *], Christopher M. Sorensen[2] and Rajan K. Chakrabarty[1, 3, †]

**Affiliations:**

[1] Center for Aerosol Science and Engineering, Department of Energy, Environmental and Chemical Engineering, Washington University in St. Louis, St. Louis, Missouri 63130, USA

[2] Department of Physics, Kansas State University, Manhattan, Kansas 66506, USA

[3] McDonnell Center for the Space Sciences, Washington University in St. Louis, St. Louis, Missouri 63130, USA

[*] These authors contributed equally to this work

[†] Corresponding author: chakrabarty@wustl.edu

**Supplementary Information**

Section I. Derivation of Equation (2)

Section II. Determination of $t_{IGP}$ from DLCA

Section III. Derivation of Equation (4)

Section IV. Determination of $t_c$ from DLCA

Section V. Derivation of Equation (10)

Section VI. The power-law growth of DLCA aggregates in the cluster-dense regime

**Section I. Derivation of Equation (2)**

The off-lattice DLCA model operates with a unit timescale $t_s$ during which monomers move by a root-mean-squared-displacement $\sqrt{\langle \delta^2 \rangle}$ that equals the monomer diameter $2a$. In three-dimensional space,

$$\langle \delta^2 \rangle = 6Dt \tag{S1}$$

where $D = k_B T/(6\pi\mu a)$ is the monomer diffusivity when under Stokes-Einstein diffusion. Substituting the $\langle \delta^2 \rangle$, $t$, and $D$ by $4a^2$, $t_s$, and $k_B T/(6\pi\mu a)$, respectively, Eq. (S1) becomes $t_s = 4\pi\mu a^3/k_B T$, which is Eq. (2) in the main text.

## Section II. Determination of $t_{IGP}$ from DLCA

The determination of $t_c$ from DLCA data follows the rule $\langle N(t_{IGP}) \rangle = N_{IGP}$. Visually, this treatment is shown in FIG. S1. Note that the characteristic time is reported in units of $t_s$.

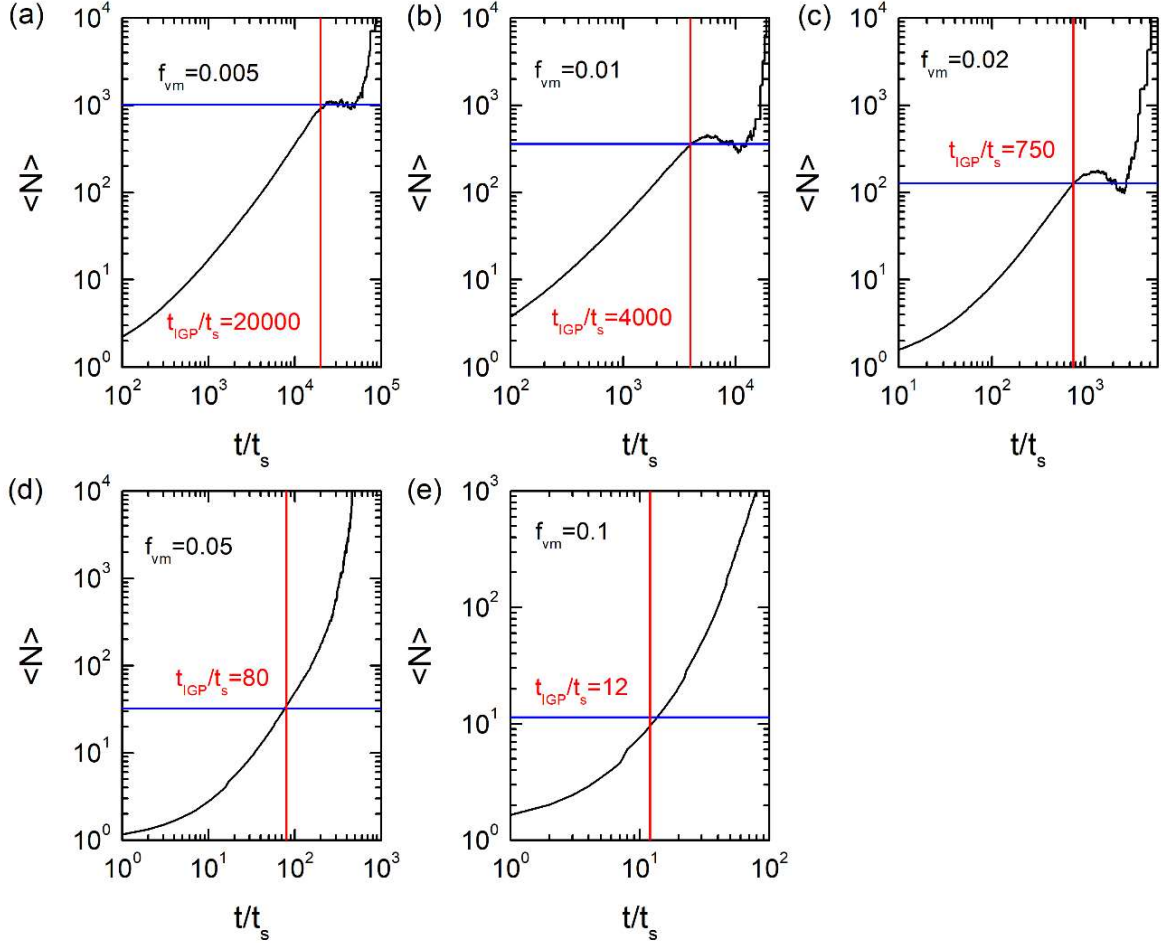

**FIG. S1. Increase in $\langle N \rangle$ as a function of simulation time $t/t_s$ for DLCA starting out with $f_{vm}$ = 0.005 (a), 0.01 (b), 0.02 (c), 0.05 (d), and 0.1 (e). Black solid curves represent $\langle N \rangle$. Blue solid lines mark the values for $N_{IGP}$ calculated per Eq. (3). Red lines mark the critical $t_{IGP}/t_s$, at which $\langle N \rangle$ reaches $N_{IGP}$.**

**Section III. Derivation of Eq. (4)**

The volume occupied by the solid components of an aggregates is $V_{agg} = \frac{4}{3}\pi a^3 N$. For the average cluster at IGP, we have,

$$V_{agg,IGP} = \frac{4}{3}\pi a^3 N_{IGP} \tag{S2}$$

Combining Eq. (3.1) and (3.2) in the main text, we get,

$$N_{IGP} = f_{vm}^{D_f/(D_f-3)} k_0^{3/(3-D_f)} \left(\frac{D_f}{D_f+2}\right)^{3D_f/(6-2D_f)} \tag{S3}$$

Substituting the $N_{IGP}$ in Eq. (S2) using (S3), we get the dimensionless form of aggregate volume,

$$\frac{V_{agg,IGP}}{a^3} = \frac{4}{3}\pi f_{vm}^{D_f/(D_f-3)} k_0^{3/(3-D_f)} \left(\frac{D_f}{D_f+2}\right)^{3D_f/(6-2D_f)} \tag{S4}$$

The volume enveloped by the MES of the aggregates is $V_{MES} = \frac{4}{3}\pi R^3$, where $R$ is the perimeter radius [1] of the cluster and it is related to $R_g$ per $R = R_g[(D_f+2)/D_f]^{1/2}$. Thus, for the MES of the average cluster at IGP, we have,

$$V_{MES,IGP} = \frac{4}{3}\pi R_{g,IGP}^3 \left(\frac{D_f+2}{D_f}\right)^{3/2} \tag{S5}$$

Substituting the $R_{g,IGP}$ in Eq. (S5) using (3.2) in the main text, we get the dimensionless form of MES volume,

$$\frac{V_{MES,IGP}}{a^3} = \frac{4}{3}\pi f_{vm}^{3/(D_f-3)} k_0^{3/(3-D_f)} \left(\frac{D_f}{D_f+2}\right)^{3D_f/(6-2D_f)} \tag{S6}$$

Combining Eq. (S4) and (S6), we get the expression for the free volume within MES,

$$\frac{V_{MES,IGP}-V_{agg,IGP}}{a^3} = \frac{4}{3}\pi(1-f_{vm})\left[k_0\left(\frac{D_f}{D_f+2}\right)^{D_f/2} f_{vm}^{-1}\right]^{3/(3-D_f)} \tag{S7}$$

The Stokes-Einstein diffusivity of the average cluster at IGP is written as $D_{ag,IGP} = \frac{k_B T}{6\pi\mu R_{g,IGP}}$, and when combined with Eq. (2) in the main text, it yields to a dimensionless expression:

$$\frac{t_s D_{agg,IGP}}{a^2} = \frac{2}{3}\left(\frac{R_{g,IGP}}{a}\right)^{-1} \tag{S8}$$

Substituting the $R_{g,IGP}$ using Eq. (3.2) in the main text, the (S8) becomes:

$$\frac{t_s D_{agg,IGP}}{a^2} = \frac{2}{3}\left[f_{vm}^{-1} k_0 \left(\frac{D_f}{D_f+2}\right)^{3/2}\right]^{1/(D_f-3)} \tag{S9}$$

The characteristic time $t_d = \frac{(V_{MES,IGP}-V_{agg,IGP})^{2/3}}{6 D_{agg,IGP}}$ was introduced per Eq. (S1). After being combined with Eq. (S7) and (S9), the expression of $t_d$ yields to the dimensionless form shown in the main text as Eq. (4):

$$\frac{t_d}{t_s} = \frac{1}{4}\left[\frac{4}{3}\pi(1-f_{vm})\right]^{2/3} \left(\frac{D_f}{D_f+2}\right)^{(2D_f+3)/(6-2D_f)} (k_0 f_{vm}^{-1})^{3/(3-D_f)}$$

## Section IV. Determination of $t_c$ from DLCA

The determination of $t_c$ from DLCA data follows the rule $n_{tot}(t_c) = n_{tot,0}/2$, as Eq. (5) of the main text suggests. Visually, this treatment is shown in FIG. S2. Note that the characteristic time is reported in units of $t_s$.

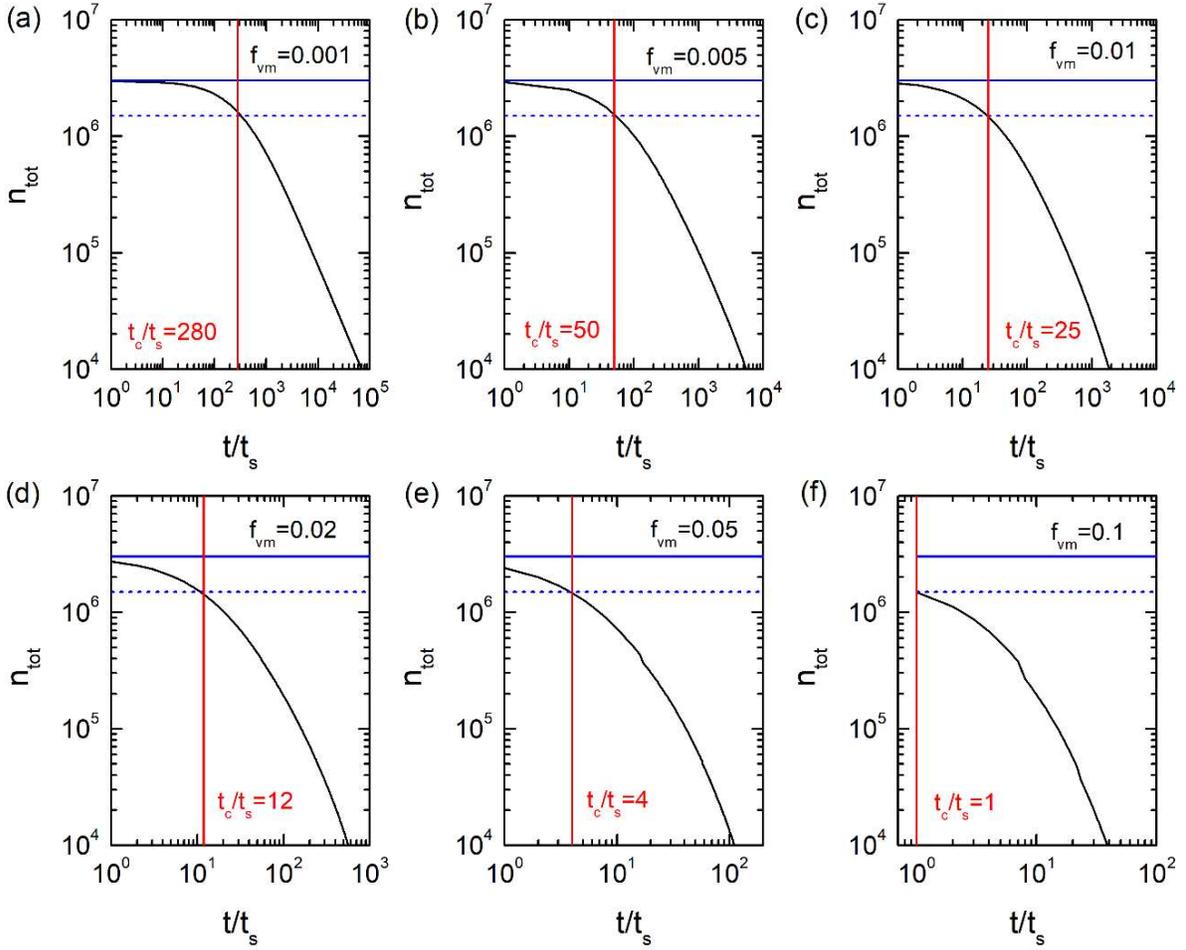

**FIG. S2.** Decrease in $n_{tot}$ as a function of simulation time $t/t_s$ for DLCA starting out with $f_{vm}$ = 0.001 (a), 0.005 (b), 0.01 (c), 0.02 (d), 0.05 (e), and 0.1 (f). Black solid curves represent $n_{tot}$. Blue solid-lines and dotted lines mark the values for $n_{tot,0}$ and $n_{tot,0}/2$, respectively. Red lines mark the critical $t_c/t_s$, at which $n_{tot}$ reaches $n_{tot,0}/2$.

## Section V. Derivation of Equation (10)

Ref [1] outlined that solution to SE with homogenous kernel yields the average DLCA cluster mass $\langle N \rangle \propto (1 + t/t_c)^z$, and we use the following power-law relationship to describe the particle growth in the cluster-dense regime:

$$\langle N \rangle = b(1 + t/t_c)^z, \tag{S10}$$

where $b$ and $z$ are assumed to take constant values for simplicity. Combining Eq. (S10) and Eq. (S3) under the condition $\langle N \rangle = N_{IGP}$ at $t = t_{IGP}$ gives

$$t_{IGP} = t_c \left[ b^{-\frac{1}{z}} f_{vm}^{\frac{D_f}{(D_f-3)z}} \left(\frac{D_f}{D_f+2}\right)^{\frac{3D_f}{(6-2D_f)z}} k_0^{\frac{3}{(3-D_f)z}} - 1 \right]. \tag{S11}$$

Replacing the $t_c$ with the right-hand-side of Eq. (9) in the main text, we get

$$\frac{t_{IGP}}{t_s} = \frac{1}{4}(f_{vm}^{-1} - \Phi_P^{-1}) \left[ b^{-\frac{1}{z}} f_{vm}^{\frac{D_f}{(D_f-3)z}} \left(\frac{D_f}{D_f+2}\right)^{\frac{3D_f}{(6-2D_f)z}} k_0^{\frac{3}{(3-D_f)z}} - 1 \right],$$ which is Eq. (10) in the main text.

**Section VI. The power-law growth of DLCA aggregates in the cluster-dense regime**

FIG. S3 shows the onset of the power-law relationship between $\langle N \rangle$ and $1 + t/t_c$ in the cluster-dense regimes. Parallel trends with a constant exponent $z$ of about 1.5 are observed for DLCA systems starting out with different values of $f_{vm}$. A Rigorous treatment requires parameterizing the prefactor $b$ in Eq. (S10) as a function of $f_{vm}$. We, however, use a constant $b = 0.2$ for a simpler final expression of Eq. (10) in the main text.

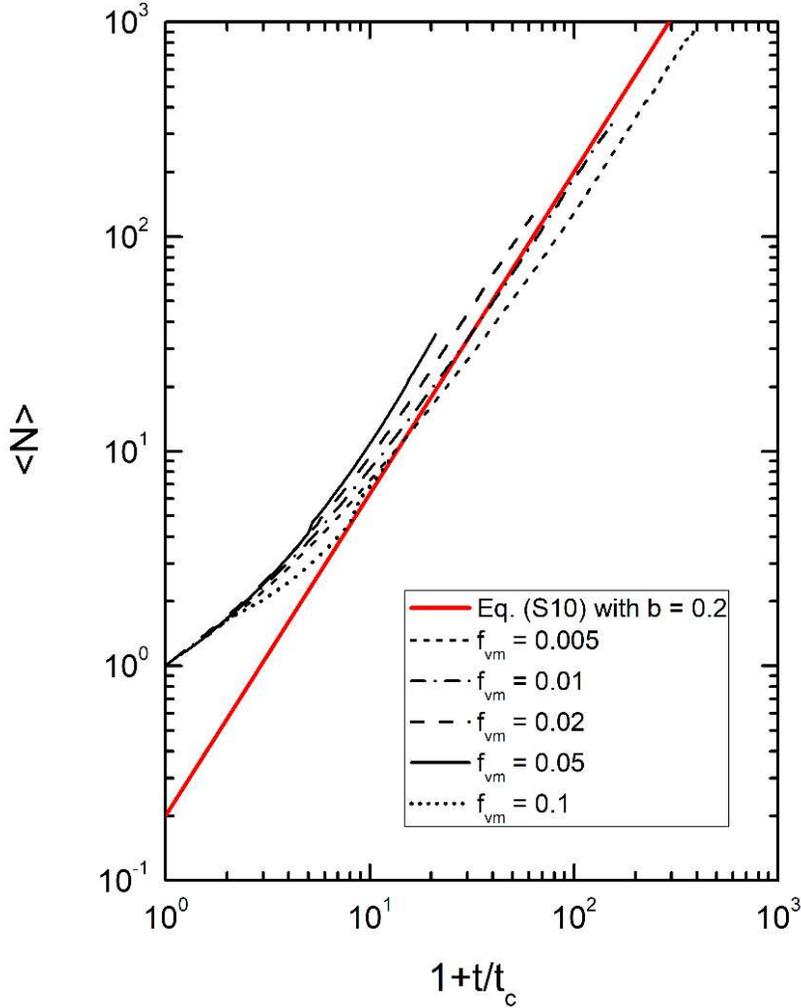

**FIG. S3. Increase in $\langle N \rangle$ as a function of $1 + t/t_c$ for the aggregates in DLCA systems during the time interval between the begin of aggregation and the IGP.**